\documentclass[twocolumn,showpacs,preprintnumbers,amsmath,amssymb]{revtex4}

\usepackage{amsmath}
\usepackage{amssymb}
\usepackage{amsfonts}
\usepackage{graphicx}

\newcommand{\ket}[1]{\lvert{#1}\rangle} 
\newcommand{\bra}[1]{\langle{#1}\rvert} 

\newcommand{\braket}[2]{\langle#1\vert#2\rangle}

\begin{document}

\title{Security of coherent state quantum cryptography in the presence of
  Gaussian noise}

\author{Matthias Heid}
\author{Norbert L\"utkenhaus}
\affiliation{Quantum Information Theory Group, Institut f\"ur Theoretische
  Physik I and Max-Planck Research Group, Institute of Optics, Information and
  Photonics, Universit\"at Erlangen-N\"urnberg, Staudtstr. 7/B2, 91058
  Erlangen, Germany\\
  Institute for Quantum Computing and Department of Physics and Astronomy, University of Waterloo, 200 University Avenue West, Waterloo, Ontario N2L 3G1, Canada}
\date{\today}

\begin{abstract}

We investigate the security against collective attacks of a continuous variable quantum key distribution scheme in the asymptotic key limit for a realistic setting. The quantum channel connecting the two honest parties is assumed to be lossy and imposes Gaussian noise on the observed quadrature distributions. Secret key rates are given for direct and reverse reconciliation schemes including postselection in the collective attack scenario. The effect of a non-ideal error correction and two-way communication in the classical post-processing step is also taken into account.
\end{abstract}

\maketitle
\section{Introduction}

The goal of quantum key distribution (QKD) is to distribute a key between two honest parties, usually called Alice and Bob, which is provably secure against any eavesdropper Eve. It is assumed that Eve is only limited by the laws of physics. From a practical point of view, implementations using coherent states as input signals and variations of homodyne \cite{grosshans03a,silberhorn02a,ralph00a,hillery00a,hirano03a,lance05a} detection seem to be promising, since they can readily be realized experimentally. Moreover, it has been suggested that homodyne detection can be performed at high repetition rates in continuous variable (CV) QKD to boost the secret key rate \cite{grosshans03a}. The security of these schemes has been investigated before \cite{grosshans03a,grosshans03b,grosshans05a,navascues05a,namiki03a,namiki04a,namiki05a,namiki06a,iblisdir04a,silberhorn02b} and unconditional security has been proven for losses of up to 1.4 dB \cite{assche05a}. Though advancing our understanding of these schemes, no analysis has been provided that would give an unconditional secure key over channels with higher losses or channels imposing excess noise on the observed quadratures. In this article we will present an analysis that derives a security result under the assumption of collective attacks and the observation of Gaussian noise. The result is derived in the infinite key limit, thus ignoring statistical effects. Though restricted in this sense, there are strong indications that these restriction can be lifted, so that our results, if successfully combined with other results will lead to the desired unconditional security. We expect this will be a fair representation of the (still missing) full unconditional security proof. We compare different techniques of extracting a secret key from shared classical data such as postselection (PS) and reverse reconciliation (RR). Moreover, our approach can be modified to include two-way communication in the classical post-processing step of the protocol and non-ideal error correction.
 
Any QKD protocol can be thought of consisting of two phases. The goal of the first phase is to distribute an effectively entangled state between Alice and Bob \cite{curty04a,curty05a}. This entanglement does not need to be present in actual physical systems. Instead, it can be brought in as a theoretical construct \cite{bennett92c,curty04a}, as explained in more detail in Sec. III. In practise, Alice and Bob will use a prepare-and-measure scheme, where Alice encodes some bit-value $i$ into non-orthogonal signal states. She sends a sequence of $n$ such states over the quantum channel to Bob. In general, Eve might interact coherently with these $n$ states. We restrict ourselves here to the case of collective attacks, where Eve attaches an independent probe to each signal. Then the total state shared between Alice and Bob will be of tensor product form $\rho_{AB}^{\otimes n}$. 

Eve, however, may keep her quantum states $\rho_{E,i}$, which summarize all her knowledge about the sent signals until the second phase of the protocol is completed. In this phase, Alice and Bob use an authenticated but otherwise insecure classical channel to correct for errors in their bit-strings and to cut out Eve's knowledge about the key (privacy amplification)\cite{bennett95a}. The information sent over the classical channel becomes available to Eve who then can optimize her collective measurements on the quantum states. For this scenario of collective attacks, we apply the generic approach by Devetak and Winter \cite{devetak05a} to give a lower bound on the secret key rate.

The security analysis presented here applies to the situation where the quantum channel connecting Alice and Bob is lossy with single-photon transmittivity $\eta$ and imposes Gaussian excess noise $\delta$ on the quadrature distributions.  This kind of noise is typically seen in the experiments \cite{lorenz04a,lorenz05a}. It has been shown that a distillation of a secret key in CV-QKD is only possible when
\begin{equation}
  \delta<2 \eta\;,
\end{equation}
because otherwise the correlations between Alice and Bob could have originated from a separable state \cite{namiki04a}. Here, the excess noise $\delta$ is determined via homodyne measurements. It can easily be verified that our calculated lower bounds on the secret key rate for the various types of protocols are well in the regime of quantum correlations.

In this article, we compare different approaches to distill a key for a CV prepare-and-measure scheme. We assume that the quantum channel between Alice and Bob can be verified to be Gaussian through tomographic complete measurements and that Eve is restricted to collective attacks. While the observation of a Gaussian channel is certainly a restriction, it should be noted that this scenario is typically encountered in practise. Moreover, recent work \cite{navascues06a,garcia06a} indicates that the Gaussian attack might be optimal for the non-postselected protocols considered here. However, it is still an open problem to relate this result to protocols including announcements and postselection. Furthermore, there is hope to find a quantum de Finetti  like argument \cite{renner05a} valid in the regime of continuous variables to extend the security against collective attacks to unconditional security, as this can already be done in finite dimensions. Since we are only interested in the key rate in the asymptotic limit, we do not consider any finite size effects in our analysis. A complete security proof would have to resolve these issues.

We consider a protocol where Alice uses coherent states as signals and send through Eve's domain to Bob, who performs a heterodyne measurement onto the received states. It is known that one can improve the secret key rate if one introduces reverse reconciliation (RR) \cite{grosshans03a,grosshans03b}. This means that Bob decides on a raw key based upon his measurement results and consequently sends Alice correction information over the public channel in the error correction step of the protocol. Another way to improve the performance of the protocol is to employ postselection (PS)\cite{silberhorn02b}: Bob only retains measurement outcomes that are closely correlated to Alice in order to gain some advantage over Eve. This approach can lead to positive secret key rates for direct reconciliation (DR) schemes beyond the so called 3dB loss limit \cite{grosshans02a}. Since both approaches are not mutual exclusive, we consider combinations of DR and RR with PS. If one takes a realistic error correction protocol into account, it has been shown that it is necessary to introduce a postselection step in the RR protocols to retrieve the initial advantage that RR has over DR \cite{heid06a}.

This article is organized as follows. In Sec. II we introduce the QKD protocol under investigation. Then we describe the state distribution scheme in an entanglement based scheme. The fact that state distribution in our protocol can be seen as Alice and Bob performing tomographic complete measurements lets us restrict Eve's knowledge about the signals. This is applied to the Gaussian noisy channel in the next section. In Sec. V we modify our protocol and let Alice and Bob partially announce their measurement outcomes. This defines independent effective binary channels. Next, we calculate a lower bound on the secret key rate for each binary channel independently. The last section contains the numerical optimized secret key rates and our conclusion.

\section{The protocol}

We consider a prepare-and-measure scheme where Alice encodes her bit value into the modulation of coherent states $\ket{\alpha}$ as signals. The complex amplitude $\alpha=\alpha_{x}+i \alpha_{y}$ is chosen at random according to a symmetric Gaussian probability distribution 
\begin{equation}\label{apriori}
  p(\alpha)=\frac{1}{2 \pi \kappa} \mathrm{e}^{-\frac{|\alpha|^{2}}{2 \kappa}}\;,
\end{equation}
centered around the origin. Alice's assigns her signal the bit value 0 (1) if the real part of the sent amplitude $\alpha_{x}$ is positive (negative). The states $\ket{\alpha}$ are then sent through Eve's domain to Bob.  Bob performs a heterodyne measurement on the received states $\rho_{B}^{\alpha}$, which is mathematically equivalent to a projection onto a coherent state $\ket{\beta}=\ket{\beta_{x}+i \beta_{y}}$. He obtains the measurement outcome $\beta$ with probability
\begin{equation}\label{Qfunct}
  p(\beta|\alpha)=\frac{1}{\pi}\bra{\beta}\rho_{B}^{\alpha}\ket{\beta}
\end{equation}
and assigns a bit value 0 (1) whenever his measurement outcome $\beta_{x}$ is positive (negative).

After Bob has measured out the received states, Alice announces partially the amplitude of the sent signals as $a=\{|\alpha_{x}|,\alpha_{y}\}$ and Bob announces respectively partially his measurement outcome as $b=\{|\beta_{x}|,\beta_{y}\}$. As we will see in Sec. V, this announcement will enable us to decompose the problem into effective independent binary channels.

\section{Replacement of the source and complete tomographic measurements}

The starting point of our analysis is that we rephrase the state preparation step in the prepare-and-measure setup in an entanglement based way. This can be done by supplying Alice with a suitable source of entangled states. One part of the entangled state is kept by Alice whereas the other part is sent through the quantum channel to Bob. This scheme is a valid description of the prepare-and-measure scheme, if a measurement performed by Alice onto her part of the entangled state effectively prepares the desired conditional state of the prepare-and-measure scheme with the proper \emph{a priori} probabilities. As we show later, this can be done for the protocol introduced in the previous section. Moreover, both measurements performed by Alice and Bob turn out to be tomographical complete in our case.

After preparing $n$ entangled states, Alice and Bob share the state $\rho_{AB}^{\otimes n}$, since we restrict Eve to collective attacks.  Without loss of generality one can assume that $\rho_{AB}$ originates from a pure three party state $\ket{\Psi_{ABE}}$. Eve holds the purifying environment $\rho_{E}$ of $\rho_{AB}$ which summarizes her knowledge about the distributed states. 

As we restrict ourselves to the case when both measurements performed by Alice and Bob are tomographic complete, they can in principle reconstruct their shared state $\rho_{AB}$. However, we skip details of the tomography in our analysis as we are only interested in evaluating the secret key rate in the asymptotic limit as $n\rightarrow \infty$. Therefore, the security analysis presented here can be considered as incomplete. The aim is to investigate what the rate one can expect to find assuming that one solves the additional steps involving the estimate of the state shared by Alice and Bob.

From the purity of state $\ket{\Psi_{ABE}}$ it follows from Schmidt's decomposition that $\rho_{AB}=\mathrm{tr}_{E} \ket{\Psi_{ABE}}\bra{\Psi_{ABE}}$ and $\rho_{E}=\mathrm{tr}_{AB} \ket{\Psi_{ABE}}\bra{\Psi_{ABE}}$ have the same eigenvalues. Eve's reduced density matrix $\rho_{E}$ is then determined up to an arbitrary unitary operation on her system by the state tomography. This in turn completely determines Eve's knowledge about the distributed signals.

In the following we apply this kind of analysis to our protocol from Sec. II in a realistic scenario.

\section{Application to the Gaussian channel}

It has been shown by Grosshans \emph{et al.} \cite{grosshans03b}  that Alice's state preparation can formally be described in an entanglement based scheme. It corresponds to a situation where Alice has a source  under her control that produces two-mode squeezed states $\ket{\xi_{AB'}}$. If Alice performs a heterodyne measurement onto her part of the state, she effectively prepares a coherent state in the $B'$ system with Gaussian \emph{a priori} probability. As the source of two-mode squeezed states is under her control, she can choose a suitable squeezing parameter so that she indeed effectively prepares coherent states with the proper \emph{a priori} probability (\ref{apriori}). The part $B'$ of the state is then passed to Bob through Eve's domain. Bob performs a heterodyne measurement on the received state. Since this measurement is tomographical complete, we can directly apply the reasoning of the last section to this specific protocol and obtain the state in Eve's hand.

For the state tomography step, it is worth noting that Alice's reduced density matrix $\rho_{A}=\mathrm{tr} \rho_{AB}$ is fixed by preparing coherent states $\ket{\alpha}$ with the \emph{a priori} probabilities given by Eq. (\ref{apriori}). One can therefore parameterize Alice's subsystem by the variance $\kappa$ of the probability distribution $p(\alpha)$. Moreover, it suffices to check the conditional states $\rho_{B}^{\alpha}$ to estimate Eve's interference with the signals. However, we do not consider arbitrary noise imposed by Eve on the conditional states, but limit our security analysis to a scenario which is typically encountered in experiments: we assume that the states Bob receives are attenuated by the loss $\eta$ in the quantum channel and the conditional probability distributions $p(\beta|\alpha)$ as given by Eq. (\ref{Qfunct}) still have Gaussian form but are broadened by a factor 
\begin{equation}\label{excessnoise}
  \delta=2\left(\frac{\Delta^2_{\mathrm{obs}} \beta_{x}}{\Delta^2_{\mathrm{vac}} \beta_{x}}-1\right) \;,
\end{equation}
the so called excess noise. Here, $\Delta^2_{\mathrm{obs}} \beta_{x}$ denotes the observed variance of the classical probability distribution (\ref{Qfunct}) and $\Delta^2_{\mathrm{vac}} \beta_{x}$ is the corresponding variance of the vacuum. We have included the factor $2$ so that our definition of the excess noise matches the one given in Ref. \cite{namiki04a} via quadrature-` measurements. The quadrature operator $\hat{x}$ is defined as $\hat{x}=\frac{1}{\sqrt{2}}(\hat{a}+\hat{a}^{\dagger})$, where $\hat{a}$ and $\hat{a}^{\dagger}$ is the photon annihilation and creation operator. As a further assumption, we suppose that the channel adds the same amount of noise in both quadratures, so that Bob effectively verifies that he receives displaced thermal states as conditional states, denoted by $\rho_{B}^{\alpha}$. Then the probability of Bob getting the measurement outcome $\beta$ conditioned on Alice sending a coherent state with amplitude $\alpha$ is given by Eq. (\ref{Qfunct}) as
\begin{equation}\label{condprobbetaalpha}
  p(\beta|\alpha)=\frac{2}{\pi (2+\delta)}\mathrm{e}^{-\frac{2|\beta-\sqrt{\eta}\alpha|^2}{2+\delta}} \;.
\end{equation}
Since Bob's subsystem can be characterized by the estimated channel parameters, the total bipartite state $\rho_{AB}$ is given by the input variance $\kappa$, the excess noise $\delta$ and the loss $\eta$. As mentioned before, the knowledge $\rho_{AB}$ determines Eve's quantum state $\rho_{E}$ up to an arbitrary unitary operation on her system when complete tomographic measurements are available. It then follows that Eve's knowledge about the signals is fixed by the set of parameters $\kappa$, $\eta$ and $\delta$. Therefore, all attacks performed by Eve give her exactly the same amount of information about the signals as long as the channel can be verified to be Gaussian. In particular, this means that attacks like the entangling cloner \cite{grosshans03b} or the amplifier attack \cite{namiki05a,namiki06a} are equivalent in this setting and Eve retains the whole purifying environment. Recent results concerning the optimality of Gaussian attacks, when the full tomographic information is not available, can be found in \cite{navascues06a,garcia06a}. 
 
Here one can pick a specific attack to construct Eve's ancilla system $\rho_{E}$, which is only restricted in the sense that the conditional states $\rho_{B}^{\alpha}$ that Bob receives are thermal states and Eve retains the whole purifying environment of $\rho_{B}^{\alpha}$. On the other hand, the joint probability distribution $p(\alpha,\beta)$ of Alice preparing an input state with amplitude $\alpha$ and Bob obtaining a measurement outcome $\beta$ is fixed by the state tomography. It follows that Eve's conditional states $\ket{\epsilon^{\alpha,\beta}}$ already contain all her knowledge about the distributed signals. These states are pure, since they can be thought of originating from a projection measurement of the pure three party state $\ket{\Psi_{ABE}}$. Equivalently, Eve's information can also summarized in the matrix of all possible overlaps $\braket{\epsilon^{\alpha,\beta}}{\epsilon^{\alpha',\beta'}}$.

We will proceed to calculate a lower bound on the secret key rate with the specified discretisation to bit-values of continuous outcomes $\beta$ and $\alpha$. It turns out that in this case Eve will effectively have to distinguish non-Gaussian states on an infinite dimensional Hilbert space to infer the bit-value. Since this is hard to solve in general, we apply an approach to define effective binary channels as we have already done in \cite{heid06a} and let Alice and Bob partially announce $\alpha$ and $\beta$. This partial knowledge will become available to Eve, who then only needs to distinguish two nonorthogonal states on a two dimensional Hilbert space, so that we can evaluate easily all related quantities.

\section{Effective binary channels}

The security analysis presented here is limited to the collective attack scenario, so that the bipartite state between Alice and Bob after $n$ uses of the quantum channel is simply $\rho_{AB}^{\otimes n}$. Consequently, Bob's measurement outcomes $\beta$ on subsequent signals are independent. Suppose now that Alice announces the modulus of the real part $|\alpha_{x}|$ and the imaginary part $\alpha_{y}$ of the prepared amplitude $\alpha=\alpha_{x}+i \alpha_{y}$. Now Bob knows that the state he receives can only originate from the two possible states $\ket{\pm|\alpha_{x}|+i\alpha_{y}}$ and that in each distributed state one bit of classical information is encoded. Each distribution of a signal between Alice and Bob corresponds to the use of an effective binary channel defined by Alice's announcement and Bob's measurement.  From Eq. (\ref{apriori}) follows that both possible input states occur with equal probability. The probability of Alice making a certain announcement $a=\{|\alpha_{x}|,\alpha_{y}\}$ can be directly calculated form Eq. (\ref{apriori}) as
\begin{align}\label{probA}
    p(a)&=p(+|\alpha_{x}|+i \alpha_{y})+p(-|\alpha_{x}|+i \alpha_{y})\nonumber\\&=2p(|\alpha_{x}|+i\alpha_{y})=\frac{1}{\pi \kappa} \mathrm{e}^{-\frac{|a|^{2}}{2 \kappa}}\; .
\end{align}
Bob performs a heterodyne measurement on the received state. The probability that he gets the measurement outcome $\beta$ after the announcement of Alice can be calculated from Eq. (\ref{condprobbetaalpha}) as
\begin{align}\label{probbetagivena}
 p(\beta|a)&=\frac{p(\beta|a,0)p(a,0)+p(\beta|a,1)p(a,1)}{p(a,0)+p(a,1)} \nonumber\\
 &=\frac{1}{2}\left(p(\beta|a,0)+p(\beta|a,1)\right)\;,
\end{align}
where we have characterized the two possible values for the amplitude $\alpha=\pm|\alpha_{x}|+i \alpha_{y}$ by the encoded bit-value $0$ or $1$ and the announcement $a$. The conditional probabilities for Bob obtaining the measurement result $\beta$ for given announcement $a$ are directly given by (\ref{condprobbetaalpha}) as
\begin{align}\label{condprob}
p(\beta|a,0)&=\frac{2}{\pi\left(2+\delta\right)}\mathrm{e}^{-2\left(\frac{\left(\beta_{x}-\sqrt{\eta}|\alpha_{x}|\right)^2+\left(\beta_{y}-\sqrt{\eta}\alpha_{y}\right)^2}{2+\delta}\right)}\\
p(\beta|a,1)&=\frac{2}{\pi\left(2+\delta\right)}\mathrm{e}^{-2\left(\frac{\left(\beta_{x}+\sqrt{\eta}|\alpha_{x}|\right)^2+\left(\beta_{y}-\sqrt{\eta}\alpha_{y}\right)^2}{2+\delta}\right)}\;.
\nonumber
\end{align}
Similar to the announcement of Alice, we let Bob record the measured $\beta$ for each signal and publicly announce $b=\{|\beta_{x}|,\beta_{y}\}$. From Eq. (\ref{probbetagivena}) follows that 
\begin{equation*}
p(+|\beta_{x}|+i\beta_{y}|a)=p(-|\beta_{x}|+i\beta_{y}|a)\;,
\end{equation*} 
so that the probability for Bob making an announcement $b$ given Alice announced $a$ is 
\begin{equation}\label{probB}
p(b|a)=2p(+|\beta_{x}|+i\beta_{y}|a).
\end{equation}
Both announcements of Alice and Bob for a given distributed state will then define one effective binary channel. Furthermore, the error probability for Bob assigning the wrong bit-value can be computed from (\ref{condprob}) as
\begin{align}
  \mathrm{e}^{+}=\frac{p(b,+|a,1)}{p(b,+|a,0)+p(b,+|a,1)}\;,
\end{align}
where we have chosen to describe Bob's measurement outcome $\beta$ by the announcement $b$ and the sign of the measured $\beta_{x}$, which corresponds to Bob's decision on a bit-value. Respectively the error probability $e^{-}$ when he obtained a negative sign for the measured $\beta_{x}$ is given by
\begin{align}
  \mathrm{e}^{-}=\frac{p(b,-|a,0)}{p(b,-|a,0)+p(b,-|a,1)}\;.
\end{align}
From Eq. (\ref{condprobbetaalpha}) follows that each effective binary channel defined by the announcements of $a$ and $b$ is symmetric in the error rate, since
\begin{align}\label{erate}
e^{+}=e^{-} \equiv e \equiv e(|\alpha_{x}|,|\beta_{x}|)=\frac{1}{1+\mathrm{e}^{\frac{8\sqrt{\eta}|\alpha_{x}||\beta_{x}|}{2+\delta}}}
\end{align}
holds.
Each distributed state between Alice and Bob with announced $a$ and $b$ therefore corresponds to the use of an effective symmetric binary channel with error rate $e$ as given by (\ref{erate}). Each information channel contributes an amount of $1-H^{\mathrm{bin}}$ to the mutual information between Alice and Bob, whereas $H^{\mathrm{bin}}$ is the entropy of a binary symmetric channel,
\begin{align}\label{binent}
 H^{\mathrm{bin}}(e)=-e\;\mathrm{log}_{2}(e)-(1-e)\;\mathrm{log}_{2}(1-e).
\end{align}
The total mutual information between Alice and Bob $I_{A:B}$ can be calculated as a sum over all effective binary channels weighted with the appropriate probabilities (\ref{probA}) and (\ref{probB}) as
\begin{align}\label{totali1}
I_{A:B}=&\int_{0}^{\infty} \mathrm{d}|\alpha_{x}| \int_{-\infty}^{\infty} \mathrm{d}\alpha_{y} \; p(a)\times  \\&\times\int_{0}^{\infty} \mathrm{d}|\beta_{x}| \int_{-\infty}^{\infty} \mathrm{d}\beta_{y}
 \;p(b|a)\left[1-\mathrm{H}^{\mathrm{bin}}\left(e\right) \right].\nonumber
\end{align}
Since the error rate $e$ only depends on the announced values of $|\beta_{x}|$ and $|\alpha_{x}|$ one can carry out parts of the integration analytically to simplify (\ref{totali1}) as
\begin{align}\label{totalinfo}
I_{A:B}=&\int_{0}^{\infty} \mathrm{d}|\alpha_{x}|  \; p(|\alpha_{x}|)\times\\&\times\int_{0}^{\infty} \mathrm{d}|\beta_{x}| \;p\left(|\beta_{x}|\big{|}|\alpha_{x}|\right)\left[1-\mathrm{H}^{\mathrm{bin}}\left(e\right) \right].\nonumber
\end{align}
 The total probability $p(|\beta_{x}|||\alpha_{x}|)$ that Bob announces a particular value $|\beta_{x}|$ for a given announcement $a$ of Alice can be derived from (\ref{probbetagivena}) and (\ref{condprob}) as
\begin{align}\label{marginal}
 p&\left(|\beta_{x}|\big{|}|\alpha_{x}|\right)=\int_{-\infty}^{\infty}\mathrm{d}\beta_{y} \; p(b|a)\\
&=\sqrt{\frac{2}{\pi\left(2+\delta\right)}}\left(e^{-\frac{2\left(|\beta_{x}|+\sqrt{\eta}|\alpha_{x}|\right)^2}{2+\delta}}+e^{-\frac{2\left(|\beta_{x}|-\sqrt{\eta}|\alpha_{x}|\right)^2}{2+\delta}}\right)\nonumber\;
\end{align}
and the probability that Alice announces $|\alpha_{x}|$ follows from (\ref{probA}) as
\begin{equation}\label{marginalalpha}
  p(|\alpha_{x}|)=\int_{-\infty}^{\infty} \mathrm{d}\alpha_{y} p(a)=\sqrt{\frac{2}{\pi \kappa}} \mathrm{e}^{-\frac{|\alpha_{x}|^{2}}{2 \kappa}}\; .
\end{equation}
We have now quantified the mutual information between Alice and Bob. As mentioned before, Eve's information about the signals is summarized in holding conditional quantum states $\ket{\epsilon^{\alpha,\beta}}$. The announced values of $a$ and $b$ give her partial information about the distributed signals. In particular, she knows the effective binary channel that has been used by Alice and Bob and the error rate $e$ of that channel. In other words, Eve knows for a given announcement of $a$ and $b$ that she holds a convex combination of the four possible states $\ket{\epsilon^{a,b}_{0,+}},\ket{\epsilon^{a,b}_{0,-}},\ket{\epsilon^{a,b}_{1,+}},\ket{\epsilon^{a,b}_{1,-}}$ in her ancilla system, where $0$ ($1$) corresponds to an encoded bit-value $0$ ($1$) by Alice and $+$ ($-$) to Bob obtaining a positive (negative) measurement outcome for $\beta_{x}$. The state $\epsilon^{a,b}$ that Eve holds for a given announcement $\{a,b\}$ can thus be written as
\begin{align}\label{rhoE}
  \epsilon^{a,b}=&\frac{1}{2}
  \left[(1-e)
    \left(
      \ket{\epsilon^{a,b}_{0,+}}\bra{\epsilon^{a,b}_{0,+}}
      +\ket{\epsilon^{a,b}_{1,-}}\bra{\epsilon^{a,b}_{1,-}}
    \right)
  \right.\nonumber\\
  &\left.
    +e\left(
      \ket{\epsilon^{a,b}_{0,-}}\bra{\epsilon^{a,b}_{0,-}}
      +\ket{\epsilon^{a,b}_{1,+}}\bra{\epsilon^{a,b}_{1,+}}
    \right)
  \right]\;.
\end{align}
The state $\epsilon^{a,b}$ can be interpreted as a uniform mixture of states corresponding to different encoded bit-values
\begin{equation}\label{rhoEDR}
  \epsilon^{a,b}=\frac{1}{2}\left(\epsilon^{a,b}_{0}+\epsilon^{a,b}_{1}\right)\;,
\end{equation}
or as a uniform mixture of states corresponding to different signs of the measured $\beta_{x}$
\begin{equation}\label{rhoERR}
  \epsilon^{a,b}=\frac{1}{2}\left(\epsilon^{a,b}_{+}+\epsilon^{a,b}_{-}\right)\;,
\end{equation}
with
\begin{align}\label{rhoEcond}
  \epsilon^{a,b}_{0}&=
  (1-e)\ket{\epsilon^{a,b}_{0,+}}\bra{\epsilon^{a,b}_{0,+}}
  +e\ket{\epsilon^{a,b}_{0,-}}\bra{\epsilon^{a,b}_{0,-}}\nonumber\\
   \epsilon^{a,b}_{1}&=
  (1-e)\ket{\epsilon^{a,b}_{1,-}}\bra{\epsilon^{a,b}_{1,-}}
  +e\ket{\epsilon^{a,b}_{1,+}}\bra{\epsilon^{a,b}_{1,+}}\nonumber\\
   \epsilon^{a,b}_{+}&=
  (1-e)\ket{\epsilon^{a,b}_{0,+}}\bra{\epsilon^{a,b}_{0,+}}
  +e\ket{\epsilon^{a,b}_{1,+}}\bra{\epsilon^{a,b}_{1,+}}\nonumber\\
   \epsilon^{a,b}_{-}&=
  (1-e)\ket{\epsilon^{a,b}_{1,-}}\bra{\epsilon^{a,b}_{1,-}}
  +e\ket{\epsilon^{a,b}_{0,-}}\bra{\epsilon^{a,b}_{0,-}}\;.
\end{align}
If Eve wants to infer the encoded bit-value, she effectively has to distinguish the states $\epsilon^{a,b}_{0}$ and $\epsilon^{a,b}_{1}$. This case is common in QKD and we refer to it as direct reconciliation (DR). As already mentioned, there exists an inequivalent way to distill a key from exchanged quantum states in CV-QKD: with the use of strict one-way communication in the classical post-processing step of the protocol, one can force Eve to infer Bob's measurement outcome rather than the encoded bit-value. This method is called reverse reconciliation and was first pointed out by Grosshans \cite{grosshans03a,grosshans03b}. For the specific protocol investigated here this means that Eve has to discriminate $\epsilon^{a,b}_{+}$ and $\epsilon^{a,b}_{-}$ for a given effective binary channel in the RR schemes.
  
\section{Lower bound on secret key rate}

The aim of this article is to compute a achievable lower bound on the secret key rate for our specified prepare-and-measure QKD using coherent states. By now we have shown that Eve's knowledge about the distributed signals, given a certain effective binary channel is used, is summarized in the quantum states $\epsilon^{a,b}_{0}$ and $\epsilon^{a,b}_{1}$ for the DR schemes or $\epsilon^{a,b}_{+}$ and $\epsilon^{a,b}_{-}$ when RR is applied. In the following, we use a result by Devetak and Winter \cite{devetak05a}, which gives a lower bound on the secret key rate as a function of the states that Eve has to distinguish. This approach is valid in the collective attack scenario and one-way classical post-processing. Then the secret key rate $G$ is bounded from below by
\begin{equation}\label{Dev/Win}
G \geq \mathrm{I}_{A:B}-\chi \; ,
\end{equation}
with $\chi$ being Holevo's quantity \cite{holevo73b}. Since we investigate a practical QKD scheme with a specified measurement setup, we have replaced the Holevo quantity between Alice and Bob in theorem (1) of \cite{devetak05a} by the classical mutual Information $\mathrm{I}_{A:B}$. The Holevo quantity $\chi$ is defined as
\begin{align}\label{Holevo}
\chi &= S(\overline{\rho})-\sum_{i=0}^{1}\mathrm{p}_{i} S(\rho_{i})\\
\overline{\rho}&= \sum_{i=0}^{1} \mathrm{p}_{i}\rho_{i}\nonumber,
\end{align}
where $S(\rho)=-\mathrm{tr}\left(\rho \mathrm{log_{2}}\rho\right)$ denotes the von Neumann entropy and the $\rho_{i}$ are the states that Eve needs to distinguish. The announcements of $a$ and $b$ divide the state distribution into independent binary channels. It follows that we can apply the bound (\ref{Holevo}) to each effective binary channel defined by the announcement of $a$ and $b$ separately. The contribution to the mutual information between Alice and Bob per use of an effective binary channel is $1-H^{\mathrm{bin}}(e)$, where the binary entropy $H^{\mathrm{bin}}(e)$ is given by Eq. (\ref{binent}). An upper bound for Eve's information about the signals for a given announcement can be written according to Eq. (\ref{Holevo}) as
\begin{equation}\label{HolevoDR}
\chi^{a,b}_{DR} = S(\epsilon^{a,b})-\frac{1}{2}\left[S\left(\epsilon^{a,b}_{0}\right)+ S\left(\epsilon^{a,b}_{1}\right)\right]\;,
\end{equation}
when the key bit is determined by Alice's encoding procedure as in the DR schemes or as
\begin{equation}\label{HolevoRR}
\chi^{a,b}_{RR} = S(\epsilon^{a,b})-\frac{1}{2}\left[S\left(\epsilon^{a,b}_{+}\right)+ S\left(\epsilon^{a,b}_{-}\right)\right]\;,
\end{equation}
when a RR scheme is applied. We have used that the \emph{a priori} probabilities $p_{i}=\frac{1}{2}$ in a given effective binary channel for both RR and DR, as can be seen from Eqs. (\ref{rhoEDR}) and (\ref{rhoERR}). Hence we have to calculate the von Neumann entropies of the states defined in Eqs. (\ref{rhoE}) and (\ref{rhoEcond}) to bound Eve's knowledge about the key.
A lower bound can then be obtained with the help of Eqs. (\ref{probA}), (\ref{probB}) and (\ref{binent}) by summing over all independent effective binary channels as
\begin{align}\label{keytotal}
  G \geq& \int_{0}^{\infty} \mathrm{d}|\alpha_{x}| \int_{-\infty}^{\infty} \mathrm{d}\alpha_{y} \; p(a)  \int_{0}^{\infty} \mathrm{d}|\beta_{x}| \times\\ & \times\int_{-\infty}^{\infty} \mathrm{d}\beta_{y}
 \;p(b|a)\left\{\left[1-\mathrm{H}^{\mathrm{bin}}\left(e\right) \right]-\chi^{a,b}\right\}\nonumber\;,
\end{align}
where the Holevo quantity $\chi^{a,b}$ is given by Eq. (\ref{HolevoDR}) in the DR schemes and by Eq. (\ref{HolevoRR}) in the RR case. 
In the following we will explicitly calculate the Holevo quantities for these two types of protocols for a lossy and noisy Gaussian quantum channel.

\section{Eve's Information}

We have pointed out that all collective attacks that Eve might perform on the distributed signals are unitarily equivalent if the quantum channel between Alice and Bob can be verified as being symmetric and Gaussian, assuming that Eve retains the whole purifying environment. It is convenient to pick a specific attack with these properties to estimate Eve's knowledge about the distributed signals.
Here, we have chosen the entangling cloner attack \cite{grosshans03b} to carry out the calculation. In this attack, Eve taps off the signals sent by Alice with a beam-splitter and feeds one half of a two mode squeezed state in unused port of the beam-splitter. In doing so, the signals become attenuated according to the transmittivity of the beam-splitter and she introduces Gaussian excess noise on Bob's side. The amount of squeezing she uses in preparing her two mode squeezed state relates to the excess noise seen by Bob. More specifically, the state shared between Eve and Bob conditioned on Alice sending a coherent state $\ket{\alpha}$ can be constructed via
\begin{align}\label{cloner}
  \ket{\Psi^{\alpha}_{B,E}}=\hat{R}_{B,E_1}(\eta)\hat{S}_{E_1,E_2}(\xi)\ket{\alpha}_B\ket{0}_{E_1}\ket{0}_{E_2} \;,
\end{align}
with
\begin{align}
  \hat{R}_{B,E_1}(\theta)&=e^{\frac{\theta}{2}(\hat{e_1}^{\dagger}\hat{b}-\hat{b}^{\dagger}\hat{e_1})}\\
  \hat{S}_{E_1,E_2}(\xi)&=e^{-\xi \hat{e_1}^{\dagger}\hat{e_2}^{\dagger}+ \xi^{*}\hat{e_2}\hat{e_1}}\;, \nonumber
\end{align}
whereas $E_1,E_2$ label the modes in Eve's hand, $\hat{S}_{E_1,E_2}(\xi)$ denotes the two-mode squeezing operator with squeezing parameter $\xi=r \mathrm{e}^{i \phi}$ as can be found, for example, in Ref. \cite{barnett97a}. The unitary $\hat{R}_{B,E_1}(\theta)$ is associated to a beam-splitter with transmittivity $\eta$ via the identification $\sqrt{\eta}=\mathrm{cos}(\frac{\theta}{2})$. The operators $\hat{b},\hat{e_1}$ and $\hat{e_2}$ are the bosonic anihilation operators associated with the modes $E_1,E_2$ and $B$. From this, one can calculate Bob's received states by tracing out Eve's subsystem. It is easy to see that from Bob's point of view Eve effectively injects a thermal state in the beam-splitter so that Bob will observe Gaussian noise. The amount of excess noise $\delta$ is related to the squeezing parameter $\xi=r \mathrm{e}^{i \gamma}$ as $\delta=2\; \mathrm{sinh^2 r}(1-\eta)$.

From Eq. (\ref{cloner}), one can calculate Eve's states $\ket{\epsilon^{\alpha,\beta}}$ conditioned on Alice sending a coherent state $\ket{\alpha}$ and Bob obtaining the measurement outcome $\beta$ by projecting $\ket{\Psi^{\alpha}_{B,E}}$ onto $\ket{\beta}$. As before, we relabel the state $\ket{\epsilon^{\alpha,\beta}}$ in terms of the announcement $\{a,b\}$, the encoded bit-value $i\in \{0,1\}$ and the sign of Bob's measurement outcome $k\in \{+,-\}$ as $\ket{\epsilon^{a,b}_{i,k}}$. Since Eve's system is fixed up to an arbitrary global unitary on her system by the tomography step, it is sufficient to calculate the matrix of all possible overlaps $\braket{\epsilon^{a,b}_{i,k}}{\epsilon^{a,b}_{j,l}}$ to estimate Eve's knowledge. It turns out that the overlaps can be written as
\begin{align}\label{overlapmat}
&\braket{\epsilon^{a,b}_{i,k}}{\epsilon^{a,b}_{j,l}}=\\
&=\left(
\begin{array}{cccc}
  1&B e^{-i \phi}&A e^{i \psi}&AB e^{i \psi- i\phi}\\
  B e^{i \phi}&1&AB e^{i \psi+ i\phi}&Ae^{i \psi}\\
  A e^{-i \psi}&AB e^{-i \psi- i\phi}&1&B B e^{-i \phi}\\
  AB e^{-i \psi+ i\phi}&A e^{-i \psi}&B e^{i \phi}&1
\end{array}
\right)\nonumber \\
\end{align}
with 
\begin{align}
A&=\mathrm{e}^{-\left(\alpha_{x}^2\left(1-\frac{\eta}{1+\delta}\right)\right)}\\
B&=\mathrm{e}^{-\left(\beta_{x}^2\frac{\delta}{1+\delta}\right)}\;.\nonumber
\end{align}
One can get rid of the phase factors depending on $\phi$ and $\psi$ by multiplying the states $\ket{\epsilon^{a,b}_{i,k}}$ by appropriate phase factors. This is possible, since we are only interested in the construction of states of the form
\begin{equation}
  \rho=\sum_{i,k} p(i,k) \ket{\epsilon^{a,b}_{i,k}}\bra{\epsilon^{a,b}_{i,k}}\;,
\end{equation}
as can be seen from Eqs. (\ref{rhoE}) and (\ref{rhoEcond}). The states $\rho$ are obviously invariant under this transformation.

The matrix of overlaps (\ref{overlapmat}) is then of the form
\begin{align}\label{overlapmatrix}
\braket{\epsilon^{a,b}_{i,k}}{\epsilon^{a,b}_{j,l}}&=\left(
\begin{array}{cccc}1& B&A&AB\\
  B&1&AB&A\\
  A&AB&1&B\\
  AB&A&B&1
\end{array}
\right)\nonumber \\
&=\left(\begin{array}{cc}1& A\\
  A&1\\
\end{array}
\right)\otimes \left(
\begin{array}{cc}1& B\\
  B&1\\
\end{array}\right) \;.
\end{align}
From that it follows that one can write the states $\ket{\epsilon^{a,b}_{i,k}}$ as
\begin{align}\label{tensor}
\ket{\epsilon^{a,b}_{i,k}}=\ket{\epsilon^{a,b}_{i}}\ket{\epsilon^{a,b}_{k}}
\end{align}
with 
\begin{align}\label{AB}
\braket{\epsilon^{a,b}_{0}}{\epsilon^{a,b}_{1}}&=A=\mathrm{e}^{-\left(\alpha_{x}^2\left(1-\frac{\eta}{1+\delta}\right)\right)}\\
\braket{\epsilon^{a,b}_{+}}{\epsilon^{a,b}_{-}}&=B=\mathrm{e}^{-\left(\beta_{x}^2\frac{\delta}{1+\delta}\right)},\nonumber
\end{align}
where we already replaced the squeezing parameter $\xi$ by the excess noise $\delta$ observed by Bob.

Since the states under investigation can be written as a product (\ref{tensor}) of two states in two dimensional Hilbert spaces, one can expand them as
\begin{align}\label{symbasis1}
  \ket{\epsilon^{a,b}_{0}}&=c_{0}\ket{\Phi_{0}}+c_{1}\ket{\Phi_{1}}\\
  \ket{\epsilon^{a,b}_{1}}&=c_{0}\ket{\Phi_{0}}-c_{1}\ket{\Phi_{1}}
\end{align}
and
\begin{align}\label{symbasis2}
  \ket{\epsilon^{a,b}_{+}}&=c_{+}\ket{\Phi_{+}}+c_{-}\ket{\Phi_{-}}\\
  \ket{\epsilon^{a,b}_{-}}&=c_{+}\ket{\Phi_{+}}-c_{-}\ket{\Phi_{-}} \;,
\end{align}
where $\ket{\Phi_{0}}$ and $\ket{\Phi_{1}}$ form a set of orthonormal basis states for the Hilbert space spanned by $\ket{\epsilon^{a,b}_{0}}$ and $\ket{\epsilon^{a,b}_{1}}$. Respectively $\ket{\Phi_{+}}$ and $\ket{\Phi_{-}}$ form an orthogonal basis for the space spanned by $\ket{\epsilon^{a,b}_{+}}$ and $\ket{\epsilon^{a,b}_{-}}$. The coefficients $c_{0}$, $c_{1}$, $c_{+}$ and $c_{-}$ depend on the effective binary channel labeled by $a$ and $b$, though we suppress these indices now to simplify the notation. It is important, however, to keep in mind that we estimate Eve's knowledge about the signals for each effective channel independently.  The normalization condition reads 
\begin{align}\label{norm}
\left|c_{0}\right|^2+\left|c_{1}\right|^2=\left|c_{+}\right|^2+\left|c_{-}\right|^2=1.
\end{align}
and
\begin{align}\label{csover}
\left|c_{0}\right|^2-\left|c_{1}\right|^2&=\braket{\epsilon^{a,b}_{0}}{\epsilon^{a,b}_{1}}=A\\
\left|c_{+}\right|^2-\left|c_{-}\right|^2&=\braket{\epsilon^{a,b}_{+}}{\epsilon^{a,b}_{-}}=B\nonumber
\end{align}
is fixed by the overlaps (\ref{AB}). In this basis the state $\epsilon^{a,b}$ of Eq. (\ref{rhoE}) can be written as
\begin{widetext}
\begin{align}\label{overlaps}
  \epsilon^{a,b}&=\left(
    \begin{array}{cccc}
      |c_{0}|^{2}|c_{+}|^{2}& 0&0&(1-2e)c_{0}c_{1}^{*}c_{+}c_{-}^{*}\\
      0&|c_{0}|^{2}|c_{-}|^{2}&(1-2e)c_{0}c_{1}^{*}c_{+}c_{-}^{*}&0\\
      0&(1-2e)c_{0}^{*}c_{1}c_{+}^{*}c_{-}&|c_{1}|^{2}|c_{+}|^{2}&0\\
      (1-2e)c_{0}^{*}c_{1}c_{+}^{*}c_{-}&0&0&|c_{1}|^{2}|c_{-}|^{2}
    \end{array}
  \right)\;,
\end{align}
which has the eigenvalues
\begin{align}\label{eigrhoE}
  \lambda_{1,2}&=\frac{1}{2}\left[|c_{0}|^{2}|c_{-}|^{2}+|c_{1}|^{2}|c_{+}|^{2}\pm\sqrt{\left(|c_{0}|^{2}|c_{-}|^{2}+|c_{1}|^{2}|c_{+}|^{2}\right)^{2}-16e(1-e)|c_{0}|^{2}|c_{-}|^{2}|c_{1}|^{2}|c_{+}|^{2}}\right]\\
\lambda_{3,4}&=\frac{1}{2}\left[|c_{0}|^{2}|c_{+}|^{2}+|c_{1}|^{2}|c_{-}|^{2}\pm\sqrt{\left(|c_{0}|^{2}|c_{+}|^{2}+|c_{1}|^{2}|c_{-}|^{2}\right)^{2}-16e(1-e)|c_{0}|^{2}|c_{-}|^{2}|c_{1}|^{2}|c_{+}|^{2}}\right]  \nonumber\;,
\end{align}
\end{widetext}
so that we can calculate the first term of Eqs. (\ref{HolevoDR}) and (\ref{HolevoRR}) with the help of Eqs. (\ref{eigrhoE}),(\ref{csover}),(\ref{norm}) and (\ref{AB}) via the equation
\begin{equation}\label{entrho}
S(\epsilon^{a,b})=-\sum_{i}\lambda_{i}\mathrm{log}_{2}\lambda_{i}\;.
\end{equation}
The explicit expression is omitted here.

\subsection{Direct reconciliation}

In the DR protocols, Eve has to discriminate the states $\epsilon^{a,b}_{0}$ and $\epsilon^{a,b}_{1}$ as defined in Eqs. (\ref{rhoEcond}) in order to infer the bit-value encoded by Alice. These can be expressed in product form (\ref{tensor}) as
\begin{align}
  \epsilon^{a,b}_{0}&=\ket{\epsilon^{a,b}_{0}}\bra{\epsilon^{a,b}_{0}}\otimes\left[(1-e)\ket{\epsilon^{a,b}_{+}}\bra{\epsilon^{a,b}_{+}}+e\ket{\epsilon^{a,b}_{-}}\bra{\epsilon^{a,b}_{-}}\right]\\
  \epsilon^{a,b}_{1}&=\ket{\epsilon^{a,b}_{1}}\bra{\epsilon^{a,b}_{1}}\otimes\left[(1-e)\ket{\epsilon^{a,b}_{-}}\bra{\epsilon^{a,b}_{-}}+e\ket{\epsilon^{a,b}_{+}}\bra{\epsilon^{a,b}_{+}}\right]\nonumber\;.
\end{align}
With the help of the basis states $\ket{\Phi_{0}}$, $\ket{\Phi_{1}}$ and  $\ket{\Phi_{+}}$, $\ket{\Phi_{-}}$ these states can be written as 
\begin{widetext}
\begin{align}
\epsilon^{a,b}_{0}=\left(
\begin{array}{cc}\left|c_{0}\right|^2& c_{1}^{*}c_{0}\\
  c_{0}^{*}c_{1}&\left|c_{1}\right|^2
\end{array}
\right)\otimes\left(
\begin{array}{cc}\left|c_{+}\right|^2& \left(1-2e\right)c_{+}^{*}c_{-}\\
  \left(1-2e\right)c_{-}^{*}c_{+}&\left|c_{-}\right|^2
\end{array}\right)
\end{align}
\begin{align}
  \epsilon^{a,b}_{1}=\left(
  \begin{array}{cc}\left|c_{0}\right|^2& -c_{1}^{*}c_{0}\\
    -c_{0}^{*}c_{1}&\left|c_{1}\right|^2
  \end{array}
  \right)\otimes\left(
  \begin{array}{cc}\left|c_{+}\right|^2& -\left(1-2e\right)c_{+}^{*}c_{-}\\
    -\left(1-2e\right)c_{-}^{*}c_{+}&\left|c_{-}\right|^2
  \end{array}\right).\nonumber
\end{align}
\end{widetext}
It is easy to see that there exists a unitary $U$ with $\epsilon^{a,b}_{0}=U\epsilon^{a,b}_{1}U^{\dagger}$, so that $S(\epsilon^{a,b}_{0})=S(\epsilon^{a,b}_{1})$. The eigenvalues of the state $\epsilon^{a,b}_{0}$ can be obtained by first diagonalizing the sub-matrices and then taking the tensor product. Then $\epsilon^{a,b}_{0}$ reads, 
  \begin{align}
    \epsilon^{a,b}_{0}=\left(
      \begin{array}{cc}1& 0\\
        0&0
      \end{array}
    \right)\otimes\left(
      \begin{array}{cc}\lambda_{1}^{0}& 0\\
        0&\lambda_{2}^{0}
      \end{array}\right)
  \end{align}
in its eigenbasis. The eigenvalues $\lambda_{1,2}^{0}$ are given by
\begin{align}\label{eigDR}
  \lambda_{1,2}^{0}=\frac{1}{2}\left(1\pm\sqrt{1-16e(1-e)(\left|c_{+}\right|^2\left|c_{-}\right|^2)}\right)\;.
\end{align}
so that the entropy $S(\epsilon^{a,b}_{0})$ can be computed with the help of Eqs. (\ref{erate}), (\ref{AB}), (\ref{csover}) and (\ref{eigDR}) and Eve's knowledge about the distributed signals in the DR protocol is upper bounded by
\begin{equation}\label{chiDR}
  \chi^{a,b}_{DR}=S(\epsilon^{a,b})-S(\epsilon^{a,b}_{0})\;,
\end{equation}
where again the explicit expression is omitted.

\subsection{Reverse reconciliation}

In the RR schemes, the key bits are determined by the sign of Bob's measured $\beta_{x}$ component. Hence, Eve has to discriminate the corresponding states $\epsilon^{a,b}_{+}$ and $\epsilon^{a,b}_{-}$ (\ref{rhoEcond}) for a given effective binary channel. These can be written with the help of Eq. (\ref{tensor}) as
\begin{align}
  \epsilon^{a,b}_{+}&=\ket{\epsilon^{a,b}_{+}}\bra{\epsilon^{a,b}_{+}}\otimes\left[(1-e)\ket{\epsilon^{a,b}_{0}}\bra{\epsilon^{a,b}_{0}}+e\ket{\epsilon^{a,b}_{1}}\bra{\epsilon^{a,b}_{1}}\right]\\
  \epsilon^{a,b}_{-}&=\ket{\epsilon^{a,b}_{-}}\bra{\epsilon^{a,b}_{-}}\otimes\left[(1-e)\ket{\epsilon^{a,b}_{1}}\bra{\epsilon^{a,b}_{1}}+e\ket{\epsilon^{a,b}_{0}}\bra{\epsilon^{a,b}_{0}}\right].\nonumber
\end{align}
In the $\ket{\Phi_{0}}$, $\ket{\Phi_{1}}$ and  $\ket{\Phi_{+}}$, $\ket{\Phi_{-}}$basis, these states read
\begin{widetext}
\begin{align}
\epsilon^{a,b}_{+}=\left(
\begin{array}{cc}\left|c_{+}\right|^2& c_{-}^{*}c_{+}\\
  c_{+}^{*}c_{-}&\left|c_{+}\right|^2
\end{array}
\right)\otimes\left(
\begin{array}{cc}\left|c_{0}\right|^2& \left(1-2e\right)c_{0}^{*}c_{1}\\
  \left(1-2e\right)c_{1}^{*}c_{0}&\left|c_{1}\right|^2
\end{array}\right)
\end{align}
\begin{align}
  \epsilon^{a,b}_{-}=\left(
  \begin{array}{cc}\left|c_{+}\right|^2& -c_{-}^{*}c_{+}\\
    -c_{+}^{*}c_{-}&\left|c_{-}\right|^2
  \end{array}
  \right)\otimes\left(
  \begin{array}{cc}\left|c_{0}\right|^2& -\left(1-2e\right)c_{0}^{*}c_{1}\\
    -\left(1-2e\right)c_{1}^{*}c_{0}&\left|c_{1}\right|^2
  \end{array}\right).\nonumber
\end{align}
\end{widetext}
Similar as in the previous subsection, the states $\epsilon^{a,b}_{+}$ and  $\epsilon^{a,b}_{+}$ are unitarily equivalent, so that it suffices to calculate $S(\epsilon^{a,b}_{+})$ to determine the upper bound (\ref{HolevoRR}) of Eve's information about the signals for the RR protocols. The eigenvalues $\lambda^{+}_{1,2}$ of $\epsilon^{a,b}_{+}$ turn out to be
\begin{align}\label{lambda+}
  \lambda_{1,2}^{+}=\frac{1}{2}\left(1\pm\sqrt{1-16e(1-e)(\left|c_{0}\right|^2\left|c_{1}\right|^2)}\right)\;,
\end{align}
so that we can easily estimate Eve's knowledge about the distributed states with the help of Eqs. (\ref{lambda+})and (\ref{entrho}) as
\begin{equation}\label{chiRR}
  \chi_{RR}^{a,b}=S(\epsilon^{a,b})-S(\epsilon^{a,b}_{+})\;.
\end{equation}

\section{Secret key rate and postselection}

By now, we have calculated the individual terms of an upper bound $\chi^{a,b}$ on Eve's information about the raw key for DR  and for RR protocols, given that an effective information channel is used. We have also shown that the mutual information shared between the two honest parties per effective binary channel labeled by the announcement of $a$ and $b$ is given by $1-H^{\mathrm{bin}}$, with $H^{\mathrm{bin}}$ being the entropy of a symmetric binary channel (\ref{binent}). The total secret key rate can thus be calculated as a sum over all binary channels according to Eq. (\ref{keytotal}). Since neither the mutual information $(1-H^{\mathrm{bin}})$ between Alice and Bob nor Eve's information $\chi^{a,b}$ depend on the announced values of  $\alpha_{y}$ and $\beta_{y}$, one can simplify Eq. (\ref{keytotal}) as
\begin{align}\label{totalkey}
G&\geq\int_{0}^{\infty} \mathrm{d}|\alpha_{x}|  \; p(|\alpha_{x}|)\int_{0}^{\infty} \mathrm{d}|\beta_{x}| \;p(|\beta_{x}|\big{|}|\alpha_{x}|)\times\nonumber\\&\;\;\;\;\; \times\left[1-\mathrm{H}^{\mathrm{bin}}\left(e\right) -\chi^{a,b}\right]\nonumber\\ 
&=\int_{0}^{\infty} \mathrm{d}|\alpha_{x}|  \; p(|\alpha_{x}|)\int_{0}^{\infty} \mathrm{d}|\beta_{x}| \;p(|\beta_{x}|\big{|}|\alpha_{x}|)\Delta I(a,b)\;,
\end{align}
where the probabilities $p(|\alpha_{x}|)$ and $p(|\beta_{x}|\big{|}|\alpha_{x}|)$ are given by Eqs. (\ref{marginalalpha}) and (\ref{marginal}). 

The term $\Delta I(a,b)$ quantifies the average information theoretic advantage of Alice and Bob over Eve for a given effective channel. Since we have calculated this quantity for all channels separately, we can improve the performance of the protocols by dismissing effective channels whenever $\Delta I(a,b)$ is negative and hence Eve knows more about the distributed signals than Alice and Bob. This procedure is called postselection. Even in absence of noise, a postselection procedure is for example necessary to lead to a positive secret key rate beyond 3 dB losses for the DR protocols \cite{silberhorn02a}. For RR schemes, all effective binary channels contribute a positive amount to the secret key rate, if one only takes losses in the quantum channel into account \cite{heid06a}. In this scenario, postselecting the measurement outcomes cannot improve the secret key rate. This is however no longer true if the channel imposes excess noise $\delta$ on the signals, so that postselection can improve the performance of the RR schemes in this more general setting.

\section{Numerical results and discussion}

\begin{figure}
\includegraphics[width=0.48\textwidth]{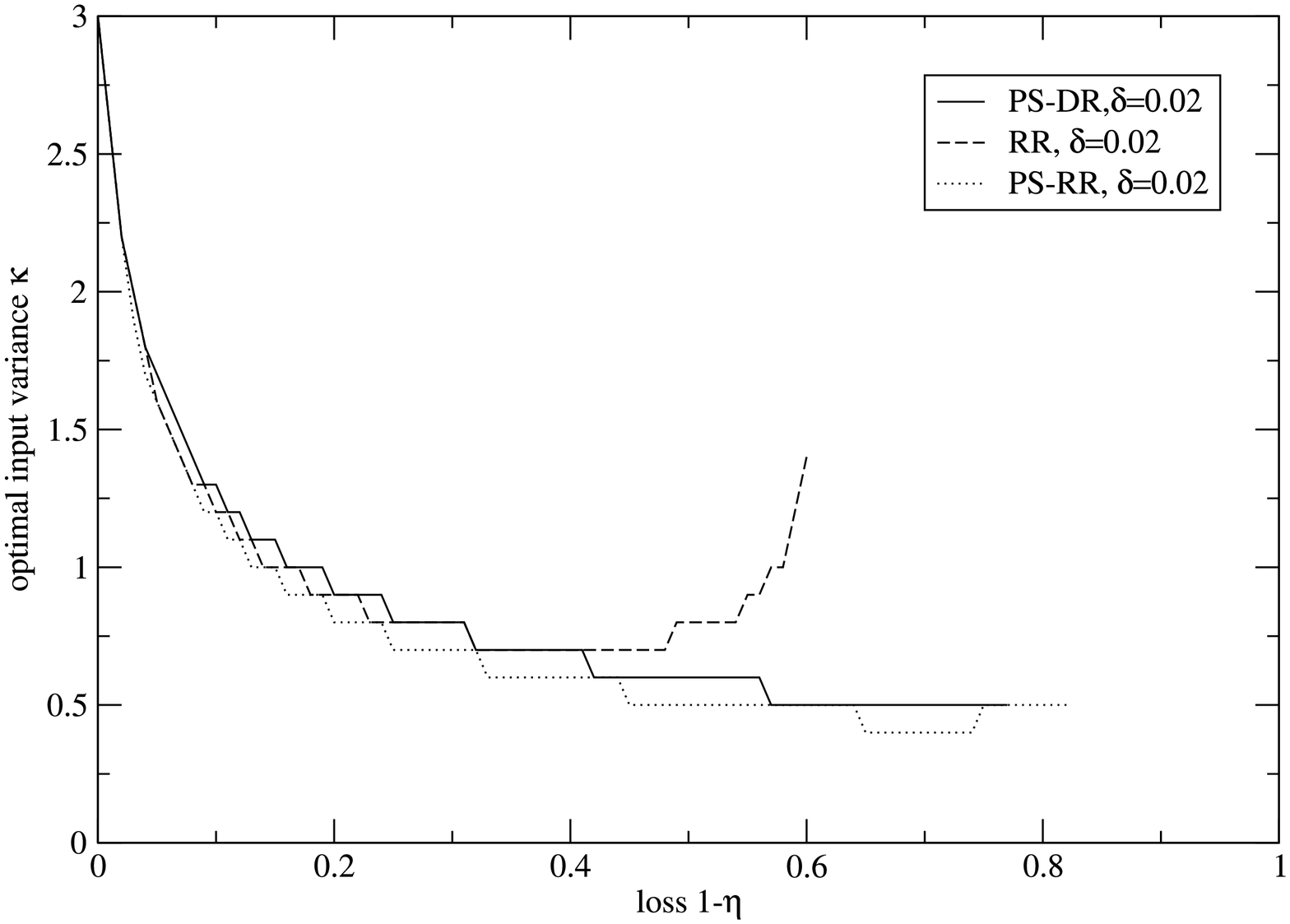}
\caption{\label{Var} Optimal values for the input variance $\kappa$ vs. loss $1-\eta$ for various protocols. All graphs shown correspond to an excess noise $\delta$ of 2\%.}
\end{figure}

Now we have everything at hand to evaluate the secret key rate $G$ numerically. For a given excess noise $\delta$ and transmission $\eta$ we can optimize the input variance $\kappa$ for best performance. Optimal values for the input variance $\kappa$ are given in Fig. (\ref{Var}). For numerical purposes, we restrict ourselves to vary the variance $\kappa$ between $0.1$ and $3$. The optimal variance $\kappa$ diverges in the limit $\eta\rightarrow 1$. Apart from that, the optimal variances fall well inside the region in which we optimize $\kappa$.

\begin{figure}
\includegraphics[width=0.48\textwidth]{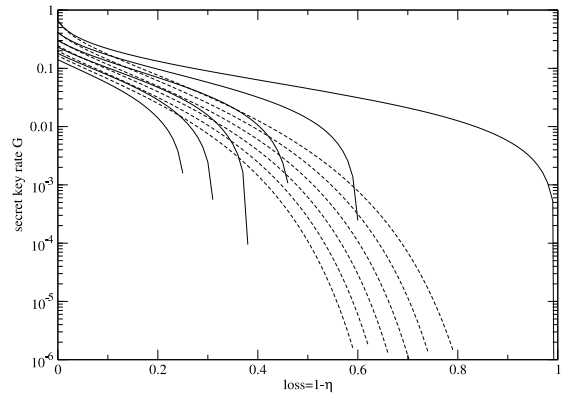}
\caption{\label{RRvsDRPS} Comparison of the secret key rate $G$ versus
  loss $1-\eta$ for the PS-DR (dashed lines) and the RR (solid lines) scheme. The secret key rates shown correspond to an excess noise $\delta$ of $\{0, 0.02, 0.04, 0.06, 0.08, 0.1\}$ and decrease with increasing excess noise.}
\end{figure}

Fig. (\ref{RRvsDRPS}) shows our results for the RR and the postselected DR scheme. 
As expected, the secret key rate $G$ decreases with increasing excess noise $\delta=\{0, 0.02, 0.04, 0.06, 0.08, 0.1\}$. However, the noise affects the non-postselected RR scheme much stronger than the postselected DR scheme (PS-DR). The RR protocol looses most of its initial advantage even for a low excess noise of 2 \%. This can be counteracted by introducing a postselection step in the RR protocols, as proposed in \cite{heid06a}.
\begin{figure}
\includegraphics[width=0.48\textwidth]{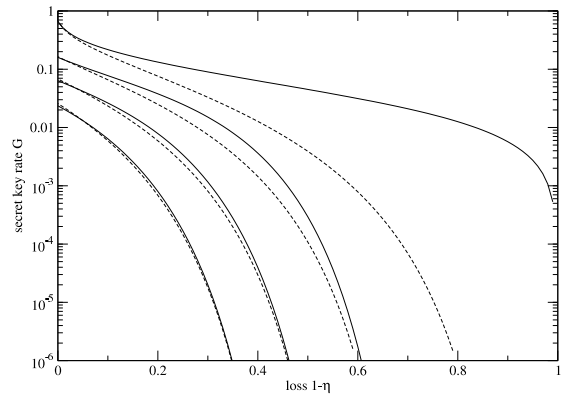}
\caption{\label{RRPSvsDRPS} Combination of postselection and reverse reconciliation. Secret key rates $G$ are plotted for the PS-DR (dashed lines) and the PS-RR (solid lines) protocols and versus the channel loss $1-\eta$. The excess noise $\delta$ varies as $\delta=\{0, 0.1, 0.2, 0.3\}$.}
\end{figure}

After introducing a postselection step in the RR scheme (PS-RR), the protocol performs more robustly against increasing excess noise $\delta$, as can be seen in Fig. (\ref{RRPSvsDRPS}). Now the PS-RR scheme performs better than the DR counterpart for all values of the excess noise, though the behavior of the secret rate gets more and more similar for increasing excess noise. This shows again that it is advantageous to combine postselection with reverse reconciliation for best performance in the presence of Gaussian noisy quantum channels.
However, here we assume that all observed excess noise occurs in the quantum channel and can therefore be exploited by Eve. As a benchmark, one can tolerate an excess noise of about $\delta=0.2$ if the quantum channel has  50\% transmittivity. It follows that the applicability of the protocols is very limited with this conservative assumption.

\subsection{Two-way communication}

The security analysis presented here assumes that the communication between Alice and Bob in the classical post-processing step is strictly one-way. From a practical point of view however, it is favorable to give lower bounds to the secret key rate $G$ for two-way communication, since these kind of protocols can easily be implemented with known error correction procedures like CASCADE. In principle, the bound (\ref{Dev/Win}) requires one-way communication to be used. This can be circumvented however, if one reveals all information to Eve that is in principle obtainable by an eavesdropper when two-way classical post-processing is used. Following earlier treatment in Ref. \cite{nl99a}, one can assume for two-way error correction the worst-case scenario in which the precise position of the errors in Bob's data become publicly known. Then it does not matter anymore, whether Alice or whether Bob make subsequent announcements. Note that in CASCADE, Bob's announcements are completely determined by the error position, and therefore no longer need to be taken into account when calculating the cost of error correction. Given this knowledge, Eve can update the state $\epsilon^{a,b}$ (\ref{rhoE}) that summarizes her knowledge about the distributed signals and the remaining communication can be chosen to be one-way. Then Eq. (\ref{Dev/Win}) is again valid, but the states $\epsilon^{a,b}$ now include this additional information. It follows that Eve either holds the state
\begin{equation}\label{epsnoerror}
  \epsilon^{a,b}_{\mathrm{no\;error}}=\frac{1}{2}\left(\ket{\epsilon^{a,b}_{0,+}}\bra{\epsilon^{a,b}_{0,+}}+\ket{\epsilon^{a,b}_{1,-}}\bra{\epsilon^{a,b}_{1,-}}\right)
\end{equation} 
or
\begin{equation}
  \epsilon^{a,b}_{\mathrm{error}}=\frac{1}{2}\left(\ket{\epsilon^{a,b}_{0,-}}\bra{\epsilon^{a,b}_{0,-}}+\ket{\epsilon^{a,b}_{1,+}}\bra{\epsilon^{a,b}_{1,+}}\right)
\end{equation} 
in her ancilla system. Obviously, the probability that an error in the bit assignment occurs is given by $e$. It is then easy to show that Eve's information about the signals for a given announcement $(a,b)$ is bounded by
\begin{align}\label{chi2}
  \chi_{\mathrm{2-way}}^{a,b}&=e\chi_{\mathrm{error}}^{a,b} + (1-e)\chi_{\mathrm{no\;error}}^{a,b}\\
&= e S\left(\epsilon^{a,b}_{\mathrm{error}}\right)+(1-e) S\left(\epsilon^{a,b}_{\mathrm{no\;error}}\right)\;,\nonumber
\end{align}
whereas the second line follows from the fact that here Eve has to distinguish pure states. Furthermore, since $ \epsilon^{a,b}_{\mathrm{no\;error}}$ and  $\epsilon^{a,b}_{\mathrm{error}}$ are unitarily equivalent, $S\left(\epsilon^{a,b}_{\mathrm{error}}\right)= S\left(\epsilon^{a,b}_{\mathrm{no\;error}}\right)$ and 
\begin{equation}\label{chi2way}
  \chi_{\mathrm{2-way}}^{a,b}=S\left(\epsilon^{a,b}_{\mathrm{error}}\right)=S\left(\epsilon^{a,b}_{\mathrm{no\;error}}\right)\;.
\end{equation}
The entropy $S\left(\epsilon^{a,b}_{\mathrm{no\;error}}\right)$ is given by the eigenvalues $\lambda^{2-way}_{1,2}$ of $\epsilon^{a,b}_{\mathrm{no\;error}}$ (\ref{epsnoerror}). It is straight forward to show that these are given by
\begin{align}
  \lambda^{2-way}_{1}&=|c_{0}|^2|c_{-}|^2+|c_{1}|^2|c_{+}|^2\\
  \lambda^{2-way}_{2}&=|c_{0}|^2|c_{+}|^2+|c_{1}|^2|c_{-}|^2\nonumber
\end{align}
with $|c_{0}|^2,|c_{-}|^2,|c_{1}|^2,|c_{+}|^2$ implicitly given by Eqs.(\ref{norm}) and (\ref{csover}).
\begin{figure}
\includegraphics[width=0.48\textwidth]{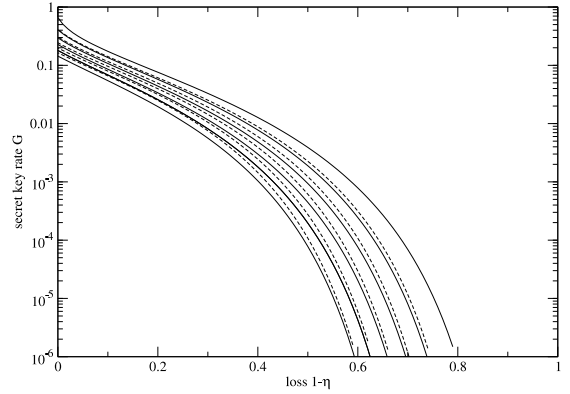}
\caption{\label{twoway}Secret key rates $G$ for postselected protocols and two-way communication (solid lines) in comparison to the one-way PS-DR protocol (dashed lines). The excess noise $\delta$ varies between $0$ and $0.1$ as in Fig. (\ref{RRvsDRPS}). For $\delta=0$, the curve for two-way communication coincides with the one for the PS-DR protocol.}
\end{figure}

Figure (\ref{twoway}) shows our numerical results for the secret key rate $G$ with two-way communication in comparison to the postselected DR results. If there is no channel excess noise $\delta$ present, we recover our previous result that the DR-PS rate coincides with the two-way rate \cite{heid06a}. Moreover it can be seen that the knowledge about the error positions does not improve Eve's position significantly in our analysis. Even in the presence of excess noise $\delta$, the DR-PS rate gives a good approximation to the two-way bound. A practical implementation using two-way error correction codes like CASCADE will therefore yield a secret key rate close to the one-way DR-PS rate.

\subsection{Practical error correction}

We extend our analysis presented here to a more realistic scenario and take the effect of a non-ideal error correction procedure into account. 

\begin{figure}
\includegraphics[width=0.48\textwidth]{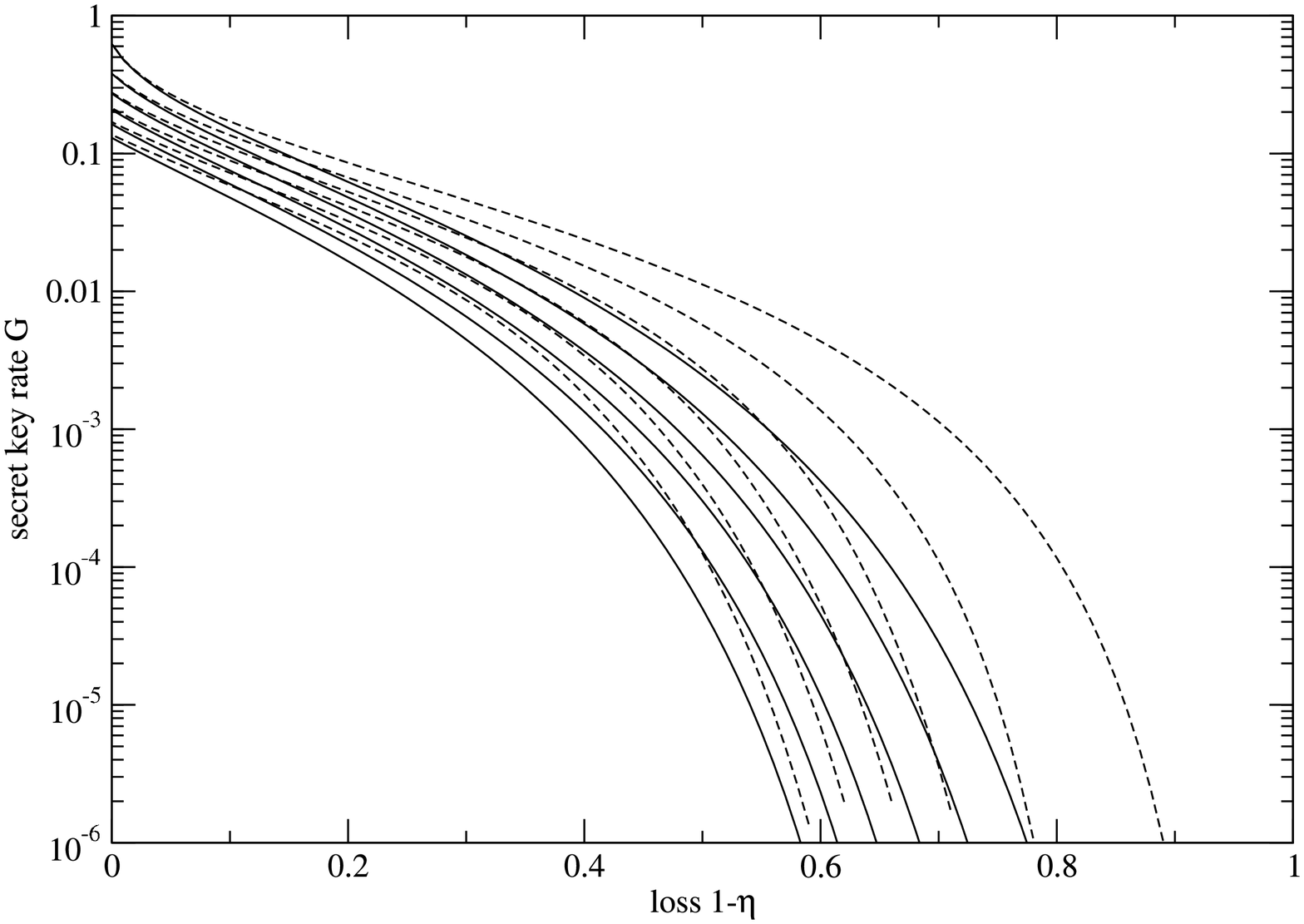}
\caption{\label{2WRRprac}Secret key rates $G$ for postselected protocols using the two-way error correction scheme CASCADE (solid lines). For comparison, key rates for the PS-RR protocol with one-way codes, that are as efficient as CASCADE are also shown (dashed lines). The excess noise $\delta$ varies between $0$ and $0.1$ as in Fig. (\ref{RRvsDRPS}).}
\end{figure}

The key rate (\ref{totalkey}) gives the theoretical achievable key rate if a perfect error correction procedure is available. In practise however, error correction codes that work exactly at this so called Shannon limit \cite{shannon48a} are not known. Realistic error correction codes, like CASCADE \cite{brassard93a} work close to that limit. This can be included by modifying Eq. (\ref{totalkey}) as
\begin{align}\label{totalkeyprac}
G&\geq\int_{0}^{\infty} \mathrm{d}|\alpha_{x}|  \; p(|\alpha_{x}|)\int_{0}^{\infty} \mathrm{d}|\beta_{x}| \;p(|\beta_{x}|\big{|}|\alpha_{x}|)\times \nonumber\\&\times\left[1-f(e)\mathrm{H}^{\mathrm{bin}}\left(e\right) -\chi^{a,b}\right] \;,
\end{align}
where the function $f(e)$ represents the efficiency of the error correction procedure and is a function of the error rate $e$. As a benchmark, we assume that the used error correction is as efficient as CASCADE. For our numerical evaluation, we therefore use a linear fit to the values given in Table \ref{effcascade}. 
\begin{table}
    \begin{center}
      \begin{tabular}{c|c}
	$e$ & $f(e)$\\
	\hline \hline
	0.01 & 1.16\\
	0.05 & 1.16\\
	0.1 & 1.22\\
	0.15 & 1.32\\
      \end{tabular}
      \caption{\label{effcascade} Efficiency of Cascade \cite{brassard93a} for different values of the error rate $e$}
    \end{center}
  \end{table}
For two-way communication, Eve's knowledge $\chi^{a,b}$ in Eq. (\ref{totalkeyprac}) is given by Eq. (\ref{chi2way}). Following this approach, we can give secret key rates which are attainable with todays technology. Numerical results are shown in Fig. (\ref{2WRRprac}).

Reverse reconciliation clearly requires one-way communication. On the other hand, developing practical and efficient one-way codes is still work in progress. It is therefore interesting to see how much secret key rate one would gain if one applies a one-way code that is as efficient as CASCADE. This can easily computed via Eq. (\ref{totalkeyprac}) whereas $\chi^{a,b}$ is given by Eq. (\ref{chiDR}) for the DR protocol or by Eq. (\ref{chiRR}) for the RR scheme. 

Fig. (\ref{2WRRprac}) shows also a comparison between two-way protocols and PS-RR. The error correction procedure is assumed to have the same efficiency as CASCADE. It can be seen that one-way PS-RR has a significant advantage over the attainable two-way protocol only for very low values of the channel excess noise $\delta$. This indicates that the development of efficient one-way codes, as currently under investigation by several groups, will significantly benefit RR protocols if the channel excess noise can be assumed to be of the order of a few percent.

\section{Conclusion}
 
In conclusion, we have addressed security issues for a CV-QKD scheme in a practical setting. It is important to include a postselection procedure in both the RR and DR schemes to ensure that the protocols perform robust against Gaussian excess noise. 

We have shown that a implementation using two-way error correction yields a secret key rate close to the rate of the one-way direct reconciled protocol. As the excess noise increases, the secret key rates for the one-way direct or reverse reconciled protocols become more and maore similar to the ones obtainable by two-way communication. Finally, we compute the secret key rate for a protocol that is readily implementable using the error correction code CASCADE.

\section{Acknowledgements}

We thank Frederic Grosshans for helpful discussions. This work has been supported by the EU-IST network SECOQC, and the German Research Council (DFG) under the Emmy-Noether program.

\bibliographystyle{apsrev}

\end{document}